\documentclass[10pt,a4paper,twocolumn,nofootinbib,floatfix,showpacs,aps]{revtex4-1}
\usepackage{amsmath,amssymb,bm}
\usepackage[utf8]{inputenc}
\usepackage{array}
\usepackage{pgfplots}
\usepackage{slashed}
\usepackage{subeqnarray}
\usepackage{amsfonts}
\usepackage{color}
\usepackage{graphicx}
\usepackage[dvips]{epsfig}
\usepackage[english]{babel}
\usepackage{bm}
\usepackage{hyperref}
\usepackage{graphicx}
\usepackage{epstopdf}
\usepackage{tensor}
\usepackage{indentfirst} 

\def\met {{\not\!\! E_T}}

\begin{document}
\title{Constraining Elko Dark Matter at the LHC with Monophoton Events}
\author{Alexandre Alves}
\email{aalves@unifesp.br}
\author{M. Dias}
\email{marco.dias@unifesp.br}
\affiliation{Departamento de F\'isica, Universidade Federal de S\~ao Paulo (UNIFESP)\\
Diadema-SP-Brazil}

\author{F. de Campos}
\email{fernando.carvalho@ict.unesp.br}
\affiliation{Departamento de Engenharia Ambiental, ICT-Universidade Estadual Paulista (UNESP),\\
S\~ao Jos\'e dos Campos-SP-Brazil}
\author{L. Duarte}
\email{laura.duarte@feg.unesp.br}
\author{J. M. Hoff da Silva}
\email{hoff@feg.unesp.br}
\affiliation{Departamento de F\'isica e Qu\'imica, Universidade Estadual Paulista (UNESP),\\
 Guaratinguet\'a-SP-Brazil}


\begin{abstract}
A mass dimension one fermion, also known as Elko, constitutes a dark matter candidate which might interact with photons at the tree level in a specific fashion. In this work, we investigate the constraints imposed by unitarity and LHC data on this type of interactions using the search for new physics in monophoton events. We found that Elkos which can explain the dark matter relic abundance mainly through electromagnetic interactions are excluded at the 95\%CL by the 8 TeV LHC data for masses up to 1 TeV.
\end{abstract}

\maketitle
\noindent

\section{Introduction}

The so-called Elko field, a set of four spinors which carry dual helicity, are constructed to be candidates to dark matter coming from the quantum field theory realm. These spinors are eingenspinors of the charge conjugation operator, $C$, with positive eigenvalues (self-conjugated) and negative eigenvalues (anti-self-conjugated). Moreover, as a trace of its own construction, they have canonical mass dimension one. That mismatch of the mass dimension of Elko
and Standard Model (SM) fermions restricts its interactions with SM particles \cite{elko}\cite{elko2}. Another striking feature of Elko is contained in their spin sums. In this structure it appears a term violating full Lorentz symmetry. However, in the context of Very Special Relativity (VSR), this problem can be bypassed and one obtain sums of spin invariant by any $HOM(2)$ Lorentz subgroup transformation \cite{vsr1, vsr}. Recently, a subtle deformation in the dual structure of Elko spinor has been proposed \cite{2017}. Within this new formulation all properties inherent to the Elko field are maintained, and the impasse with Lorentz violation seems solved.  

Returning to our main point, the mass dimension of the Elko field allows unsuppressed gauge-invariant tree level couplings with photons, Higgs boson pairs and self-interactions~\cite{elko}. Until now, only the interaction with the Higgs bosons had been phenomenologically investigated~\cite{elko,Alves:2014kta,Alves:2014qua}. The Higgs portal-type interactions help the Elkos to evade several experimental constraints, however, its electromagnetic interaction is likely to strongly constrain the model by collider, direct and indirect detection experiments of dark matter searches. 

Monophoton events are one of the promising search channels for new physics analysis \cite{1}\cite{2}. The missing energy, $\met$,
associated with the production of neutral stable particles on colliders is an important requirement to characterize dark matter, whose identity is currently unknown. Ordinary fermions, which can only couple to photons through dimension-7 operators like $\frac{1}{\Lambda^3}\bar{\psi}\psi F_{\mu\nu}F^{\mu\nu}$, have been searched for at the LHC in monophoton events. The latest bound comes from the 13 TeV run~\cite{Sirunyan:2017ewk} and amounts to $\Lambda\gtrsim 600(400)$ GeV for dark matter of 1(1000) GeV. 

A monophoton signal arises in the context of Elko models due its interaction with the electromagnetic tensor, given by the $U(1)_{em}$ invariant Lagrangian
\begin{eqnarray}
\mathcal{L}_{int} = g_{e}\stackrel{\neg}{\eta}(x)[\gamma^\mu,\gamma^\nu]\eta(x)F_{\mu\nu}(x)
\label{1}
\end{eqnarray}
where $g_{e}$ is a dimensionless coupling constant, with $\eta(x)$ and $\stackrel{\neg}{\eta}(x)$ denoting the quantum field for the Elko and its respective dual. 

The $g_e$ coupling is expected to be small since it contributes to the photon mass through loop corrections to the photon propagator~\cite{elko}, yet no explicit analysis have been performed to evaluate the constraints on the model concerning taking into account its dark matter candidacy. Another important theoretical requirement should be stressed -- Elko-Elko scattering via photon exchange might violate unitarity at high energies which might put another constraint to that coupling. Other experiments with potential to probe the model are, for example, the direct detection experiments like XENON and LUX, in this case directly because of the tree level electromagnetic interaction with the nucleons, and indirect data like astrophysical gamma-ray signals.

Among the various possibilities, we investigate in this work the constraints from the search of DM in monophoton events at the LHC by performing a scan over the model parameter space -- the Elko mass, $m_\lambda$, and its coupling constant with photons, $g_e$. We initially determine the region in which the model is excluded, taking into account published results for monophoton searches at the 8 TeV LHC with 20.3 fb$^{-1}$ of integrated luminosity from the ATLAS Collaboration~\cite{2}, also respecting the limit at which the interaction remains unitary in the M\"oller scattering. Our final goal is to identify the points of the parameters space which are compatible with the observed dark matter relic density and check whether they are excluded or not by the collider data. 


The paper is organized as follows: in the next Section we briefly review the Elko field, emphasizing the relevant properties for this work. Section III is devoted to obtain the Feynman rules for the field and compute a scattering process with the purpose of investigating the unitarity of coupling  at hand (\ref{1}). In Section IV we make the phenomenological analysis of model exclusion while in Section V we present the range of parameters for which Elko can be faced as dark matter from the phenomenological perspective. 

\section{A brief review on Elko field}

A formal construction of Elko provides a structure for these spinors that satisfies
\begin{align}
 C\lambda({\bf p})=\pm \lambda({\bf p})
\label{cc}
 \end{align}
where the charge conjugation operator, $C$, is written in the Weyl representation as
\begin{equation}
C = \left( \begin{array}{cc}
0 & i\Theta \\
-i\Theta & 0
\end{array} \right)K.
\label{CC}
\end{equation}
The operator $K$ is responsible for complex conjugate the spinor it is acting on and $\Theta=-i\sigma_{2}$ is the Wigner time-reversing operator for spin $1/2$ particles. 

In (\ref{cc}) we have a set of four equations, being two of them associated to eigenspinors with positive eigenvalue, the self-conjugated spinors $\lambda_{\alpha}^{S}({\bf p})$, and the remain two related to eigenspinors with negative eigenvalue, the anti-self-conjugated spinors $\lambda_{\alpha}^{A}({\bf p})$. The $\alpha$ index denotes the helicity (or type) of the spinor. The construction of the formal structure for these spinors is well developed in \cite{elko}\cite{elko2}, here we shall just pinpoint some relevant remarks to the rest of the paper. 

The structure of the dual spinor for the Elko that yields a non-null Lorentz invariant norm under boosts and rotations is given by 
\begin{equation}
\stackrel{\neg}{\lambda}^{S/A}_{\{\mp,\pm\}}( {\bf p})=\pm i[\lambda^{S/A}_{\{\pm,\mp \}}({\bf p})]^\dagger \gamma_{0}.
\label{dual}
 \end{equation}
It is observed that there is a change of helicity in the expressions of the dual. With the aid of above equations, it is possible to set down the orthonormality relations:
\begin{eqnarray}
\label{ana}\stackrel{\neg}{\lambda}^{S/A}_{\alpha}( {\bf p}){\lambda}^{S/A}_{\alpha'}( {\bf p})&=&\pm 2m\delta_{\alpha\alpha'}.
 \end{eqnarray}

The spin sums for these spinors are 
\begin{eqnarray}
\sum_{\alpha}{\lambda}^{S}_{\alpha}( {\bf p}){\stackrel{\neg}{\lambda}}^{S}_{\alpha}( {\bf p}) = m(I+\mathcal{G}(\varphi)), \\
\sum_{\alpha}{\lambda}^{A}_{\alpha}( {\bf p}){\stackrel{\neg}{\lambda}}^{A}_{\alpha}( {\bf p})= -m(I-\mathcal{G}(\varphi))
\label{sk}
 \end{eqnarray}
where $\mathcal{G}(\varphi)$ reads 
\begin{eqnarray}
\mathcal{G}(\varphi)= i\left( \begin{array}{cccc}
0 & 0 & 0 & -e^{-i\varphi}\\
0 & 0 & e^{i\varphi} & 0\\
0 & -e^{-i\varphi} & 0 & 0\\
e^{i\varphi} & 0 & 0 & 0
\end{array} \right).
\end{eqnarray}
The angle $\varphi$ is defined via the following momentum parametrization
\begin{eqnarray}
 {\bf p} = p(\sin\theta\cos\varphi,\sin\theta\sin\varphi,\cos\theta)
\end{eqnarray}
where $0\leq \theta \leq \pi$ and $0 \leq \varphi \leq 2\pi$.


As it will be necessary in further calculations, we remark that for the unusual spin sums, $\sum\limits_{\alpha}\lambda^{S}_{\alpha}({\bf p})\lambda^{S\dagger}_{\alpha}({\bf p})$, we find the following result
\begin{eqnarray}
\label{malandro} &&\sum_{\alpha}\lambda^{S}_{\alpha}({\bf p})\lambda^{S\dagger}_{\alpha}({\bf p})=E(I+\mathcal{G}(\varphi)) + \\ \nonumber 
&&p\left( \begin{array}{c c c c}
\cos\theta & e^{-i\varphi}\sin\theta &i\sin\theta & -ie^{-i\varphi}\cos\theta\\
\sin\theta e^{i\varphi} & -\cos\theta & -ie^{i\varphi}\cos\theta & -i\sin\theta\\
-i\sin\theta &ie^{-i\varphi}\cos\theta &-\cos\theta &- e^{-i\varphi}\sin\theta\\
i\cos\theta e^{i\varphi} & i\sin\theta &-\sin\theta e^{i\varphi} & \cos\theta
\end{array} \right),
\end{eqnarray} which differs from the expression presented in Ref. \cite{elko}. 

The Fourier decomposition of the Elko field may be written as 
\begin{eqnarray}\label{c1}
\eta(x) &=& \int\frac{d^{3}p}{(2\pi)^{3}}\frac{1}{\sqrt{2mE({\bf p})}}\sum_\beta[a_\beta({\bf p})\lambda_{\beta}^{S}({\bf p})e^{-ip_\mu x^\mu} \nonumber \\ 
 &+&b_{\beta}^{\dagger}({\bf p})\lambda_{\beta}^{A}({\bf p})e^{ip_\mu x^\mu}] 
\end{eqnarray}
and its dual
\begin{eqnarray}\label{c2}
\stackrel{\neg}{\eta}(x) &=& \int\frac{d^{3}p}{(2\pi)^{3}}\frac{1}{\sqrt{2mE({\bf p})}}\sum_\beta[a_\beta^\dagger({\bf p})\stackrel{\neg}{\lambda}_{\beta}^{S}({\bf p})e^{ip_\mu x^\mu} \nonumber \\ 
&+&b_{\beta}({\bf p})\stackrel{\neg}{\lambda}_{\beta}^{A}({\bf p})e^{-ip_\mu x^\mu}].
\end{eqnarray}
Here $a_\beta({\bf p})$ ($a^\dagger_\beta({\bf p})$) and $b_\beta({\bf p})$ ($b^\dagger_\beta({\bf p})$) are the annihilation and creation operators for particles and anti-particles \cite{Fr}, satisfying the Fermi-Dirac statistics
\begin{eqnarray}
\label{comu}\{a_\beta({\bf p}),a_{\beta'}^\dagger({\bf p'})\} = (2\pi)^{3}\delta^{3}({\bf p}-{\bf p'})\delta_{\beta\beta'}\\ \nonumber
\{a_\beta^\dagger({\bf p}),a_{\beta'}^\dagger({\bf p'})\} = \{a_\beta({\bf p}),a_{\beta'}({\bf p'})\}=0
\end{eqnarray}
with similar relations for $b$ operators. 

The Elko field has mass dimension one, satisfying only the Klein-Gordon equation. Hence the free Lagrangian density can be written as (with the proviso of always bearing in mind the spin sums peculiarities)
\begin{eqnarray}
 \mathcal{L}_{0}= \partial^{\mu}\stackrel{\neg}{\eta}(x)\partial_{\mu}{\eta}(x)-m^2\stackrel{\neg}{\eta}(x){\eta}(x)
\end{eqnarray}
and the perturbatively renormalizable interaction terms are
\begin{eqnarray}\label{int}
&& \mathcal{L}= h_{e}\phi^{\dagger}(x)\phi(x)\stackrel{\neg}{\eta}(x){\eta}(x)+\alpha_{e}[\stackrel{\neg}{\eta}(x){\eta}(x)]^2\\ \nonumber
&&+g_{e}\stackrel{\neg}{\eta}(x)[\gamma^\mu,\gamma^\nu]\eta(x)F_{\mu\nu}(x)
 \end{eqnarray}
where $h_{e}, \alpha_{e}$ and $g_{e}$ are dimensionless coupling constants. The first term in (\ref{int}) represents the interaction
with the Higgs field, already studied in Refs. \cite{Alves:2014kta}, \cite{Alves:2014qua},\cite{5}. The second term is the self-interaction of Elko field \cite{elko5}. The last term is the allowed interaction with the electromagnetic field, object of analysis of the present study. 
 
\section{Feynman rules and unitarity for Elko-photon coupling}

The Feynman rules for external lines of Elko in the momentum space can be read of from the contraction of field operators (\ref{c1}) and (\ref{c2}) with external particle states \cite{landau}. Hence, in order to evaluate the Feynman diagrams the following prescription will be used for the external lines:  
\begin{itemize}
\item $\frac{\lambda_{\beta'}^{S}({\bf k})}{\sqrt{m}}$ - for the $S$ particle incoming the vertex;
\item $\frac{\stackrel{\neg}{\lambda}_{\beta'}^{A}({\bf k})}{\sqrt{m}}$ - for the $A$ particle incoming the vertex;
\item $\frac{\stackrel{\neg}{\lambda}_{\beta'}^{S}({\bf k})}{\sqrt{m}}$ - for the $S$ particle outgoing the vertex;
\item $\frac{{\lambda}_{\beta'}^{A}({\bf k})}{\sqrt{m}}$ - for the $A$ particle outgoing the vertex.
\end{itemize} According to the prescription used for the quantum field operator (\ref{c1}) for the treatment of scattering amplitudes, the spinor $S$ will represent particles, whereas the $A$ spinor shall be related to the anti-particles.

The interaction vertex can be easily obtained by deriving functionally the Lagrangian of interaction (\ref{1}) with respect to the fields. Thus,
 \begin{eqnarray}
\Gamma_{\stackrel{\neg}{\lambda}\lambda A_{\mu}}=2ig_{e}[\gamma^\sigma,\slashed{q}],
\label{vint}
\end{eqnarray} where $q$ is the photon momentum. 
\begin{center}
\begin{figure}[t!]
	\includegraphics[scale=0.47]{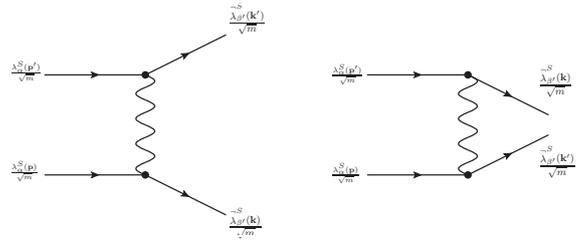}
	\caption{M\"oller scattering for Elkos with the external legs explicited.}
	\label{dm0x}
	\end{figure}
\end{center}

In order to investigate the constraints of the photon interaction from unitary arguments, we firstly calculate the M\"oller scattering for Elkos, using a \verb+GiNaC+ \cite{GiNaC} routine. The unpolarized $\eta_{\alpha}({\bf p})+\eta_{\alpha'}({\bf p'})\rightarrow \eta_{\beta}({\bf k})+\eta_{\beta'}({\bf k'})$ scattering is given by
\begin{eqnarray}
\mid\overline{\mathcal{M}}\mid ^2 = \frac{1536 g_{e}^4 (6E^4-6m^2E^2+m^4)}{m^4},
\label{final}
\end{eqnarray}
and therefore the differential cross-section in the center of mass frame for this process is
\begin{eqnarray}
 \left(\frac{d\sigma}{d\Omega}\right)_{CM}=\frac{6g_{e}^4(6E^4-6m^2E^2+m^4)}{\pi^2 E^2m^4}.
\end{eqnarray} A direct computation yields 
\begin{eqnarray}
 \sigma = \frac{24 g_{e}^4(6E^4-6m^2E^2+m^4)}{\pi E^2m^4}.
\label{final4}
 \end{eqnarray}
At high energies the cross-section grows indefinitely with energy and, as a consequence, the condition of unitarity of the $\mathcal{S}$ matrix is violated \cite{WEINBERG}. However, through the partial wave analysis it is possible to obtain the region of the parameters space in which the process remains unitary \cite{WEINBERG}\cite{unitariedade}. 

In the present case, for the M\"oller scattering the S-wave amplitude for the energies of interest is 
\begin{eqnarray}
a_{0}(\hat{s})=\frac{1}{32\pi}\int_{-1}^{1}d(\cos\theta)\mathcal{M}(\hat{s}).
\label{a}
\end{eqnarray}
To find the threshold (\ref{a}), the matrix elements were calculated using, again, a \verb+GiNaC+ \cite{GiNaC} routine. The largest amplitude for the scattering of two Elkos in the partons reference frame is
\begin{eqnarray}
 \mathcal{M}=\frac{48 g_{e}^2\hat{s}}{m^2}\; .
\end{eqnarray}

Unitarity of the scattering amplitude requires that $|\text{Re} a_{0}| \leq \frac{1}{2}$, reflecting the fact that the amplitude is bounded. Therefore the condition (\ref{a}) implies
\begin{eqnarray}
 a_{0}=\frac{3g_{e}^2 \hat{s}}{\pi m^2}\leq \frac{1}{2},
 \label{0}
\end{eqnarray} leading to the the bound 
\begin{equation}\label{uni}
\frac{g_e}{m_\lambda}\leq \sqrt{\frac{\pi}{6\hat{s}}}\; .
\end{equation}
The most stringent absolute bound for a collider search is obtained by fixing $\sqrt{\hat{s}}=\sqrt{S}$, the CM energy of the collider, in our case, 8 TeV.

\section{Relic Density Calculation}

\begin{center}
\begin{figure}[t!]
	\includegraphics[scale=0.47]{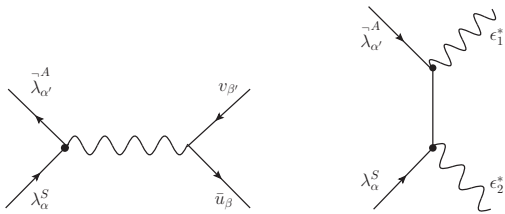}
	\caption{Annihilation of two Elkos in SM fermions and photons. They can also annihilate into $W$ bosons pairs via electromagnetic interactions.}
	\label{dm0}
	\end{figure}
\end{center}

In Fig.~(\ref{dm0}) we show two generic contributions to the Elko-Elko annihilation relevant for the determination of the relic abundance of Elkos today. The Elkos can annihilate to quarks and leptons, to photons and $W$-bosons if it is heavy enough. 
We have for the reaction \(\stackrel{\neg}{\lambda} \lambda \to \gamma \gamma\) , \[\langle \sigma_{ann}v\rangle_{\gamma\gamma} =\frac{g_e^4}{16 m\pi}\langle v^2\rangle, \]
while the reaction \(\stackrel{\neg}{\lambda} \lambda \to e^+ e^-\) leads to \[\langle \sigma_{ann}v\rangle_{e^+e^-} =\frac{g_e^2}{12 m\pi}-\frac{g_e^2}{96 m\pi}\langle v^2\rangle,\] where \(v\) is the relative velocity of the Elko particles in the annihilation process and the brackets mean that we are taking the thermal average of the quantity at hand. In our expansion $E=mv^2/\sqrt{1-v^2/4}$ since the Elko velocity is \(\frac{v}{2}\). 

We assume that there is Elko annihilation into three charged leptons and five quark flavors, with each of the latter giving a contribution of three times for the color factor and charge. Also we take account the annihilation in a top and em W bosons, given by
\[\langle \sigma_{ann}v\rangle_{W^+W^-} =\frac{(m^2-M_W^2)g_e^2e^2}{9\pi m M^2_W}+ \frac{(m^2+M_W^2)g_e^2e^2}{72\pi m M^2_W}\langle v^2\rangle\] 
Besides, we assume that at tree level \(\langle \sigma_{ann}v\rangle_{tot}=\langle \sigma_{ann}v\rangle_{e^+e^-}+\langle \sigma_{ann}v\rangle_{\gamma\gamma}+\langle \sigma_{ann}v\rangle_{W^+W^-}\).

The full procedure to determine the plane phase, shown in Fig.\ref{dm2} to our case, is described in Ref. \cite{Jungman:1995df}. It is assumed that the ``freeze-out'' temperature is \( T_f\approx \frac{m}{20}\). Since \(\langle v^2\rangle=\frac{6}{x_f}\), \(x_f=\frac{T_f}{m}\), we can express \(\langle \sigma_{ann}v\rangle_{tot}=a+6\frac{b}{x_f}\).

We numerically solve the equation
\begin{equation}x_f=\ln\left[\frac{0.0764M_{Planck}(a+6\frac{b}{x_f})c(2+c)m}{\sqrt{g_{*}(T)x_f}}\right],\label{xis1}\end{equation}
for \(x_f\), where \(g_{*}(T)\) is the number of effective relativistic degrees of freedom evaluated at \(T_f\). Besides, in (\ref{xis1}), $c=1/2$  in order to have a ten per cent error (at most), whilst $a$ and $b$ are the zeroth and first order coefficients, respectively, in the annihilation cross section of Elkos to the considered standard model particles. Now, it is possible to use \(x_f\) to calculate the present mass density as
\[\Omega_\chi=2.88\times 10^{8}/Y^{-1}_{\infty},\]
where
\[Y^{-1}_{\infty}=0.264\sqrt{g_{*}(T_f)}M_{Planck}m\left[\frac{a}{x_f}+\frac{3}{x_f^2}\left(b-\frac{a}{4}\right)\right].\]


The solid black line in Fig.~(\ref{dm2}) represents all the points where $\Omega_\chi=0.1186\pm 0.0020$ according to the  WMAP~\cite{WMAP} fit for the dark matter relic density. According to Ref.~\cite{elko}, contributions from Higgs bosons are relevant for rather light Elkos of order 10 MeV. This mass region is constrained by the unitary of M\"oller scattering amplitudes as we are going to see in the next section. Heavier Elkos, by their turn, unfortunately would require larger $h_e$ coupling to the Higgs bosons. However, if the perturbation method to the model could be guaranteed, then it might be possible to fit the DM relic abundance with smaller $g_e$ couplings. This possibility is postponed for a future investigation.

\section{Constraining the Model using the LHC data}

In order to simulate the monophoton events at the LHC, we implemented the model in \verb+Madgraph5+~\cite{madgraph} using the \verb+FeynRules+ \cite{feyn} package. We have also modified the \verb+Source/DHELAS/aloha_functions.f+ by the inclusion of the Elko field~\cite{aloha}. For this modification we parametrized the Elko field in Cartesian coordinates. 

The \verb+CheckMate+ program \cite{Drees:2013wra} was used to verify, for a given set of coupling constants and masses, whether the model is excluded or not at 95\% C.L. by comparing the result with the experimental analysis~\cite{2}. The cuts implemented by \verb+CheckMate+ to select the signals with missing energy and one identified photon were established from the ATLAS detector results. These cuts require
\begin{align} \label{cuts}
&& \met > 150\hbox{ GeV}, p_{T}^\gamma > 125\hbox{ GeV}\nonumber \\
&& |\eta^\gamma|<1.37, \Delta R(\met,\gamma)>0.4  \nonumber\\
&& \hbox{veto electrons with: } p_{T}^e > 7\hbox{ GeV}, |\eta^e|<2.47 \nonumber \\ 
&& \hbox{veto muons with: }p_{T}^\mu > 6\hbox{ GeV}, |\eta^\mu|<2.5 \hspace{2pt}. 
\end{align}
Our simulations were performed at the parton-level only. As we are going see, the conclusions shall hardly change by taking detector effects and hadronization into account.

\begin{center}
\begin{figure}[t!]
	\includegraphics[scale=0.45]{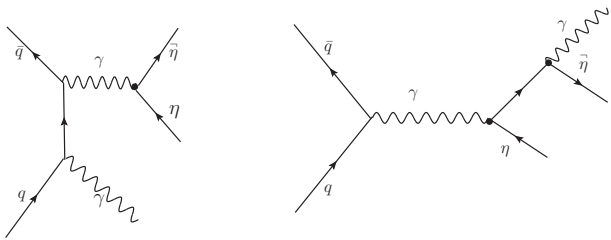}
	\caption{Elko production in monophoton events. Notice that in terms of the coupling constant, $g_e$, the graph at the right is of order $g_e^2$, but the the dominant contribution, at the left, is of order $g_e$.}
	\label{dm1}
	\end{figure}
\end{center}

The Feynman graphs for the monophoton channel are depicted in Fig.\ref{dm1}. The dominant contribution is the one where the photon is emitted from the initial state quark line, whereas the subdominant one is proportional to $g_e^2$. We took both contributions into account in our collider simulations.

\begin{center}
\begin{figure}[t!]
\includegraphics[scale=.6]{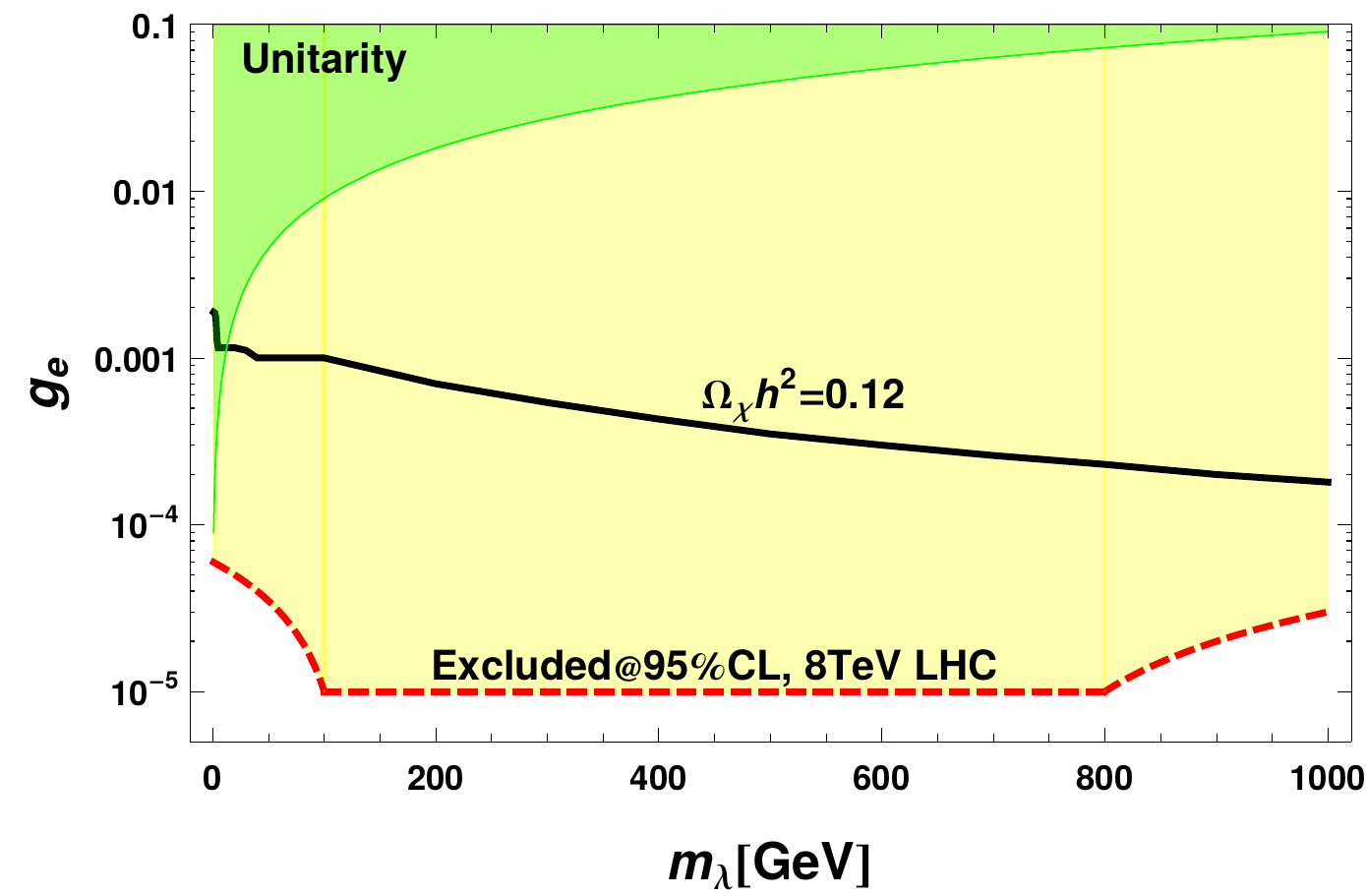}
\caption{The constraints on the Elko-photon coupling for a range of Elko masses, $m_\lambda$, and coupling constants, $g_e$. The yellow shaded region is the 95\%CL exclusion region from the 8 TeV LHC data from monophoton events. The green shaded region corresponds to the points where the unitary of M\"oller scattering of Elkos is violated. The black solid represents the points compatible with the dark matter relic density.}
\label{dm2}
\end{figure}
\end{center}

Using the ATLAS constraints for these events \ref{cuts}, the 95\%CL limit imposed on the coupling $g_{e}$, as a function of Elko mass $m_{\lambda}$, with \verb+CheckMATE+ is shown in the Fig.~(\ref{dm2}). The main background to this process is $Z\gamma$, where the $Z$-boson decays to neutrino pairs. The yellow shaded region delimited by the red dashed line represents the excluded region by the collider data. The green shaded region above the solid blue line is the region of the parameters which violate the unitary of the M\"oller scattering for Elkos in 8 TeV collisions, Eq.~(\ref{uni}) with $\sqrt{S}=8$ TeV. This is the most conservative unitarity constraint once the majority of this type of process events at the LHC would present a much lower energy and momentum transfer. Note that, even if the collider constraints were relaxed for light Elkos, the unitarity constraint would exclude them.
The main result now follows from Fig.~(\ref{dm2}) -- Elkos interacting electromagnetically and that explain the observed dark matter relic abundance are excluded by the 8 TeV LHC data and unitary of the Elko-Elko scattering. The exclusion region can actually be larger if we had taken the 13 TeV LHC data, however, the data from the 8 TeV runs suffice for our aims.

\section{Final remarks}
Elko fields are theoretically well motivated dark matter candidates due their suppressed couplings to almost the entire SM spectrum. A remarkable exception is that Elkos can interact with photons at the tree level with strength parametrized by a dimensionless coupling $g_e$. This coupling enables Elkos to be produced in monophoton events at the LHC and explain the observed dark matter relic abundance.

In this work, we show that Elkos as heavy as 1 TeV with couplings compatible with the observed dark matter relic abundance are excluded by the search for monophoton events at the 8 TeV LHC. Very light Elkos which could evade the collider constraints are, by their turn, excluded by demanding that $\stackrel{\neg}{\lambda}\lambda\rightarrow \stackrel{\neg}{\lambda}\lambda$ M\"oller scattering remains unitary up to 8 TeV.

More stringent constraints might arise either from monojet searches or using the 13 TeV LHC data already available. One way out of these hard bounds would be embedding Elkos in a multi-component dark matter model where the constraints on $g_e$ could be relaxed, for example, taking axions into account. It is also possible to increase the Higgs boson contribution to the relic density abundance, but respecting the perturbativity of the Higgs coupling. Another possibility is hypothesizing a new broken $U(1)_X$ symmetry to furnace a heavy gauge boson to intermediate the Elko couplings to the SM sector. Anyway, if the electromagnetic interactions are the dominant contributions to the relic density formation, then our results show that $g_e \gtrsim 10^{-5}$ are excluded by the LHC data for Elko masses up to 1 TeV.

\begin{acknowledgements}
The authors acknowledge the S\~ao Paulo State University (UNESP) Center for Scientific Computing (NCC/GridUNESP). JMHS thanks to Conselho Nacional de Desenvolvimento Cient\'ifico e Tecnol\'ogico-CNPq (grants  304629/2015-4; 445385/2014-6) for financial support. A. Alves acknowledges financial support from Funda\c{c}\~ao de Amparo \`a Pesquisa do Estado de S\~ao Paulo-FAPESP (grant 2013/22079-8) and CNPq (grant 307098/2014-1).
\end{acknowledgements}

\end{document}